# Relaxation of Microwave Nonlinearity in a Cuprate Superconducting Resonator

Richard A. Huizen, Sean L. Hamilton, G.T. Lenters, and S.K. Remillard, *Associate Member, IEEE*

*Abstract*— **The 2nd and 3rd order nonlinear microwave response of a superconducting YBa$_2$Cu$_3$O$_7$ thin film resonator was synchronously measured using three input tones. This technique permits the local measurement, and hence mapping, of intermodulation distortion (IMD) inside the resonator. 2nd and 3rd order IMD measured with a fixed probe relaxed in remarkably different ways after the removal of a static magnetic field. The 2nd order IMD relaxed by two different magnetic processes, a fast process that appears related to bulk remanent magnetization, and a slow process that fits the description of Bean and Livingston. The 3rd order IMD relaxes by only one process which is distinct from the two processes controlling 2nd order relaxation.**

## I. INTRODUCTION

Signal distortion in superconducting microwave resonators is usually measured as a weighted average over the entire device. Efforts to develop a theory of material nonlinearity using these global measurements are thwarted by the spatially distributed current, and hence by the spatially distributed nonlinearity inside the device under test. The interpretation of the effect of film patterning on nonlinear performance will require local measurement of nonlinearity, which will ultimately support either nonlocal electrodynamics based on perturbation of the constitutive relation [1] or local electrodynamics based on a nonlinearity threshold current [2].

A new method to measure locally generated 2nd order and 3rd order intermodulation distortion (IMD) has been

Manuscript received September 1, 2016; This material is based in part upon work supported by the National Science Foundation under Grant Numbers DMR- 1206149 and DMR- 1505617. Any opinions, findings, and conclusions or recommendations expressed in this material are those of the authors and do not necessarily reflect the views of the National Science Foundation.

*(Corresponding author: S.K. Remillard.)*

R.A. Huizen is with Grand Valley State University, Allendale, MI, 49401 USA. (e-mail: huizenr@mail.gvsu.edu).

S.L. Hamilton was with Grand Valley State University, Allendale, MI, 49401 USA. (e-mail: seanhamilton@jrauto.com).

G.T. Lenters is with Grand Valley State University, Allendale, MI, 49401 USA. (e-mail: lentersg@gvsu.edu).

S.K. Remillard is with Hope College, Holland, MI, 49423 USA. (e-mail: remillard@hope.edu).

developed [3], and in this paper its specific application to spatially resolved vortex dynamics is under investigation. Global measurements described in [4] and [5] of the 2nd and 3rd harmonic emission from MgB$_2$, including the relaxation of the 2nd harmonic in a ramping magnetic field, showed the presence of a fast process and a slow process in the 2nd harmonic relaxation, uncovering the separate impact of flux movement over the surface barrier, as well as other relaxation phenomena, on even and odd order nonlinearity. This short paper introduces the use of three-tone mixing to generate local IMD in YBa$_2$Cu$_3$O$_7$ and to produce simultaneously measured 2nd and 3rd order nonlinearity in the resonator's passband, both at essentially the same frequency. A static magnetic field is then applied leading to distinct responses of even and odd order nonlinearity to vortex matter in a critical state.

## II. LOCAL MEASUREMENT OF INTERMODULATION DISTORTION

IMD is measured in superconducting transmission line resonators by introducing multiple frequencies. The production of spurious frequencies occurs when the device is driven with a current $I$ at high frequency and the voltage responds nonlinearly as $V = I \cdot Z(I)$, where the impedance $Z$ of the superconducting transmission line depends on the current. Feeding current into the device at frequencies $f_1$ and $f_2$ produces harmonics, as well as mixing at sum and difference frequencies, which can be predicted by a Taylor expansion of $V = I \cdot Z(I)$ [6]. The 2nd order term of the two-frequency Taylor series expansion includes frequencies $2f_1$, $2f_2$, $f_2 + f_1$, and $f_2 - f_1$. The third order term includes frequencies $3f_1$, $3f_2$, $2f_2 - f_1$, $2f_1 - f_2$, $2f_2 + f_1$, and $2f_1 + f_2$. Because each order's magnitude is determined by the slope in the Taylor series expansion, it should be expected that all tones in each order have a common physical origin, as has indeed been established [7]. In this work, the decay of nonlinearity upon removal of a static magnetic field is used to investigate the different roles played by fluxons in the generation of 2nd and 3rd order nonlinearity. This paper illustrates how even and odd order nonlinearity in superconductors have different physical origins, and how IMD can be used to investigate those origins.

The circuit in Fig. 1a has three input signals at $f_1$, $f_2$, and $f_3$, providing a complicated, but more useful, host of mixing tones. 3rd order IMD occurs at, for example, $f_3 \pm (f_2 - f_1)$ and 2nd order at $f_3 \pm f_1$ or $f_3 \pm f_2$ in this *3-tone IMD* measurement [8]. $f_3$ is in-band with an adjustable input power which in these measurements was left at 0 dBm. By choosing $f_1$ and $f_2$ to be smaller than the bandwidth of the resonator, for example hundreds of KHz, both the 2nd order and the 3rd order IMD will be in the resonator's passband, and current induced at $f_1$ and $f_2$ will be localized to the region near the probe, providing a nonlinear response that has been relevant for behavioral modelling of superconducting resonators [9]. Introducing the signal at frequency $f_3$ by a separate port into the device makes this low-level IMD measurement extremely robust by using the device itself for signal source isolation.

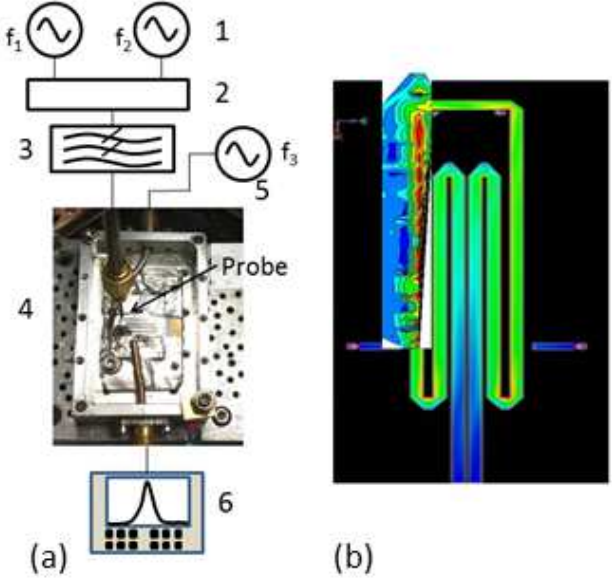

Fig. 1. (a) The setup used to measure IMD. 1. Signal generator for probing tones; 2. High isolation resistive BNC Power combiner; 3. 15 MHz Low-pass filter; 4. HTS resonator (DUT); 5. Swept signal generator for driving tone; 6. Keysight CXA spectrum analyzer. (b) $3^{rd}$ order IMD Raster scan of a portion of the resonator design used in this work. The scan is shown on top of the current distribution computed using IE3D.

The finite element simulation using HFSS [10] in Fig. 2 shows the surface current density, $K$, induced using $f_1=1$ MHz by a magnetic dipole coupling probe with a triangular tip. Because this probe frequency is outside the resonant band, the induced current remains in the vicinity of the probe, so that the IMD is also generated in the vicinity of the probe, making this a near field technique. This is thus similar to the near-field scanning microscope used elsewhere [11]-[12], which when combined with an ultrahigh resolution probe is particularly useful for identifying nanoscale defects in superconductors [13]. Measuring IMD instead of harmonics presents some unique opportunities, but reflects the same physics. Both the 2nd and the 3rd order nonlinearities are at nearly the same frequency as well as inside the resonance band of the structure, aligning the current distribution at the nonlinear frequency components with that of the quasi-TEM resonant mode [3]. Practically synchronous IMD also eliminates issues brought up by other authors including ambiguity from different time scales of the nonlinearity orders [14]-[15] and different skin depth dispersion for each order [14].

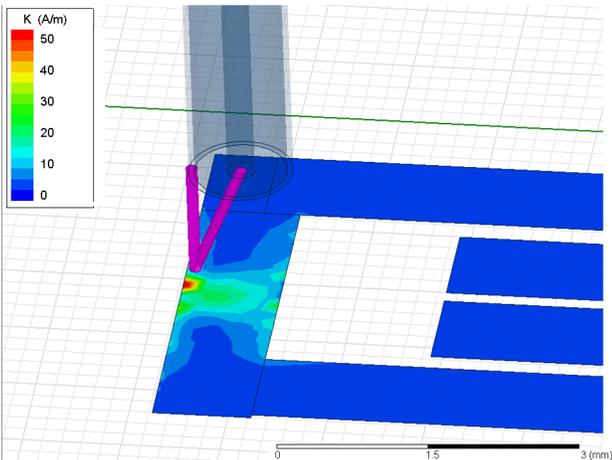

Fig. 2. Surface current is induced in the near-field of the raster probe (pink in the color version) when it is directly above a transmission line.

Raster scanning the probe across the patterned superconductor can produce spatial mapping of the even and the odd order IMD [16]. A spatial map of the 3rd orders IMD in a region of the resonator with 200 µm wide lines is shown in Fig. 1b overlaying the simulated current distribution done using IE3D [17]. In this experiment the probe was fixed at a point in the region of this scan. A static magnetic field was then applied perpendicularly to the film surface, revealing distinctions between the orders, and hence the unique role played by magnetic fluxons.

## III. IMD RELAXATION OF A CUPRATE THIN FILM

Argon ion beam milling was used to make a meandering microstrip resonator, identical to that used in [18], from a 400 nm thick $YBa_2Cu_3O_7$ (YBCO) film on a $LaAlO_3$ substrate with 200 µm wide lines, and a YBCO ground plane that was gold coated and then indium soldered to a gold plated titanium carrier. The structure shown in Fig. 1b, resonant at 840 MHz with a $T_c$ of 88.9 K and an unloaded Q of 2,500 at 77 K, was cooled with liquid nitrogen in a nonmagnetic cryostat filled with helium exchange gas and regulated with a Lakeshore 330 temperature controller.

Because of the high temperature enhancement of the nonlinear Meissner effect (NLME) [11], IMD of cuprate superconductors rises with temperature as $T_c$ is approached from below and reaches a maximum value at about 2 Kelvin below $T_c$ for YBCO, and then falls below the noise sensitivity of the instruments as the superconductor reaches $T_c$ as shown in Fig. 3 for the sample used here. This was reported to occur in the 2nd and the 3rd order IMD [18], with a peak width at half maximum for the 3rd order IMD of about 1.5 Kelvin, implying that the grains are small (less than about 500 nm) and grain boundary vortices can contribute to the nonlinearity up to a higher temperature than they do for large grained films [19]. However, in the temperature range of this experiment, $0.92T_c$ to $T_C$, the intrinsic NLME still dominates the nonlinearity.

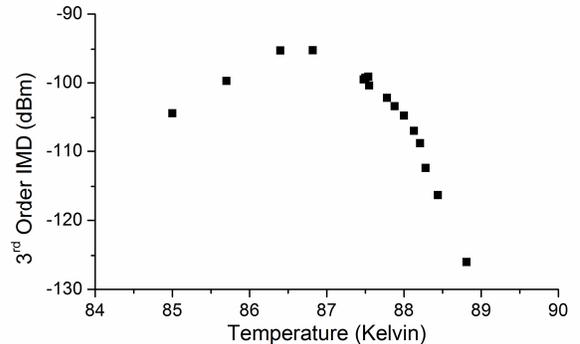

Fig. 3. The NLME peak of the 3rd order IMD in the present sample.

After Earth's-field cooling, the film was exposed to a perpendicular static magnetic field of 230 G at each temperature set-point for 10 or more minutes, after which the static field was removed and then the 2nd and 3rd order IMD were monitored for 20 or more minutes. Upon removal of the external static magnetic field, both orders of IMD relax, as seen in Fig. 4, to a new value over a period of less than 20 minutes. Relaxation did not occur during application of the field. The relaxation curve for the 3rd order IMD is concave up at all examined temperatures. The concavity of the 2nd order relaxation changes at 87.3 K. Thus here the temperature 87.3 K is referred to as an *inflection temperature* because of the change in concavity of the relaxation in the 2nd order IMD. The time after removal of the magnetic field reveals two decay processes in the 2nd order IMD. A short time constant, or *fast* process, is concave up at all temperatures. The relaxation then proceeds into a *slow* process with a long time constant. It is the slow process which undergoes inflection.

Although the 2nd order IMD in picowatts (pW) can be fit by two exponential decays $y(t) = y_0 + Ae^{-t/t1} + Be^{-t/t2}$, the only physical basis for an exponential slow process is thermally assisted flux flow [20], which is unlikely to produce signal distortion near $T_c$ especially with no applied field. Mathematically, exponential and logarithmic decays are difficult to distinguish. However, the slow process is more likely described physically by a logarithmic relaxation consistent with the Bean-Livingston theory as used with microwave nonlinearity in [4], making a more suitable fit function

$$y(t) = y_0 + A_1 e^{-t/t_1} + A_2 \log_{10}(t/t_2). \qquad (1)$$

$t_1$ is the decay constant for the fast process. $t_2$ roughly represents the time at which the slow process begins to dominate. The relaxation coefficient $A_2$ indicates the extent of relaxation in the slow process. This also requires one less free parameter than the double exponential since $y_0-A_2\log_{10}(t_2)$ is a single constant.

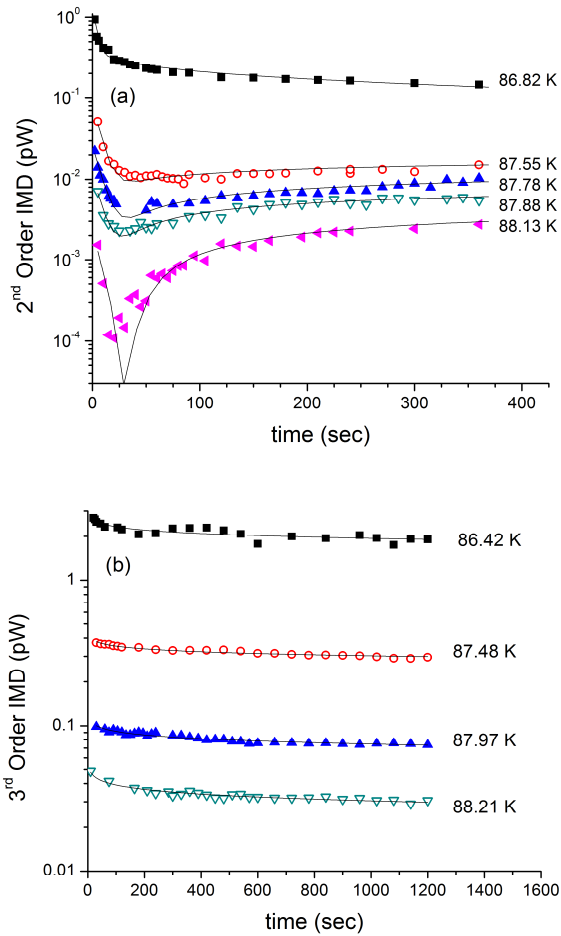

Fig. 4. (a) The 2nd order IMD and (b) 3rd order IMD decay after a 230 G static field was applied and then removed. Time equals zero when the static field was removed.

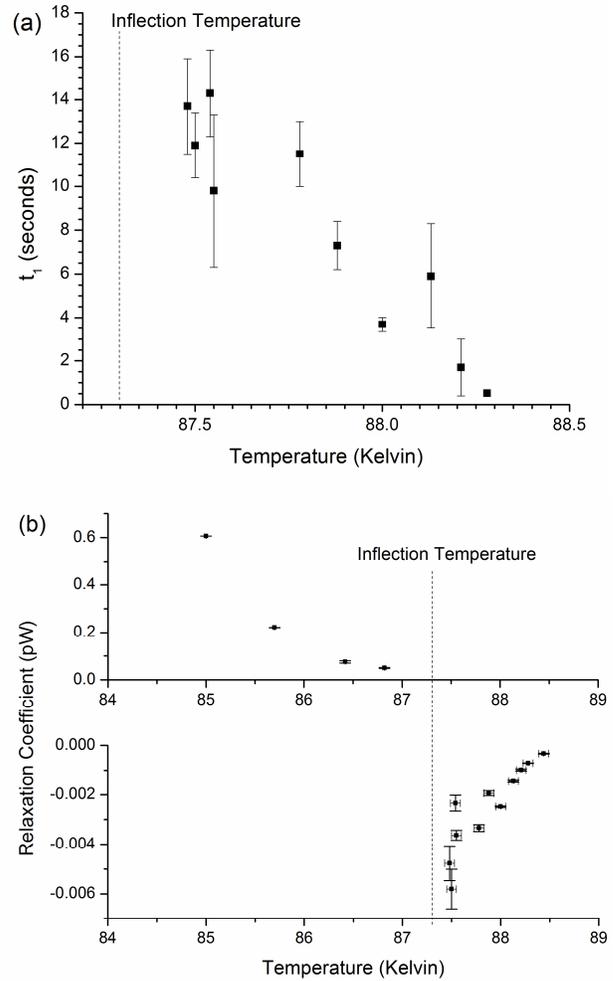

Fig. 5. (a) The time constant $t_1$ for the fast process and (b) the relaxation coefficient $A_2$ for the slow process of the 2nd order IMD. $t_1$ rapidly drops to zero above the *inflection temperature* of 87.3 K. Note the change in sign and scale of the relaxation coefficient at 87.3 K. The uncertainties are standard deviations resulting from the fits.

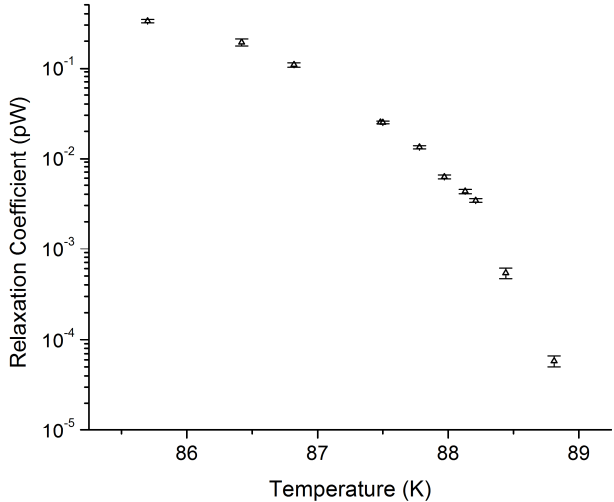

Fig. 6. The absolute values of the relaxation coefficient $A_2$ for the 3rd order IMD. The coefficients were negative, but plotted as positive so that a logarithmic scale could be shown.

The decay characteristics over the temperature range tested are shown in Fig. 5 and Fig. 6. The 2nd order fast process has a time constant in the range of 0.5-15 seconds. The slow process plays out over a time frame of 50-5,000 seconds, finishing more quickly closer to $T_C$. Above 87.3 K, the fast process time constant is strongly temperature dependent, decreasing with temperature as $T_c$ is approached from below, and vanishing above 88.3 K, 0.6 K below $T_C$. The temperature dependence of the fast process is not yet well characterized below 87.3 K. Below 87.3 K the slow process flips concavity which is apparent as a sign change in the relaxation coefficient $A_2$. At 87.3 K the slow process does not occur since the relaxation coefficient passes through zero at that temperature. Only one very weak process, with no concavity reversal, is seen in the 3rd order IMD in Fig. 4, exhibiting much less change upon relaxation than the 2nd order IMD. The 3rd order IMD in pW, fits a single weak relaxation process

$$y(t) = y_0 + A_2 \log_{10}(t/t_2) \qquad (2)$$

at each temperature with the relaxation coefficient $A_2$ shown in Fig. 5b. The relaxation coefficient becomes strongly temperature dependent above about 85 K, dropping in an exponential manner to zero at $T_C$. This is more a reflection of the strong temperature dependence of IMD than of anything else. The key point here is that, unlike the 2nd order IMD, the relaxation coefficient for 3rd order IMD does not undergo a sign change near $T_C$.

## IV. DISCUSSION

Magnetic irreversibility, which causes slow relaxation in the magnetization, can be revealed by hysteresis [20]. In experiments with MgB$_2$ using a ramped field, hysteresis in the 2nd harmonic was found to correlate to the occurrence of relaxation [4]. At all temperatures below 88.5 K reported here, hysteresis was seen in both orders of IMD. It is argued here that the entire experiment, except the highest temperature of 88.81 K, which had no hysteresis, took place below the irreversibility line [21] ruling out irreversibility cross-over as an explanation for the behavior change at 87.3 K. After the static field is discontinued, a new lower density of fluxons is energetically preferred. Fluxons exit the superconductor over a period on the order of seconds in a fast process that is likely due to remanent magnetic relaxation as some fluxons rapidly escape the surface [22]. After reaching a new equilibrium fluxon density, fluxons will still be prone to exit the superconductor by thermal activation over a surface barrier resulting in a controlled reduction in the attempt frequency [23]-[24]. This is proposed to be the slow process. The fast process is extinguished within 0.6 K of $T_c$ where the irreversibility line has possibly been crossed.

Only a single slow relaxation process is evident in the 3rd order IMD. In stark contrast to the slow process of the 2nd order IMD, the 3rd order IMD decays with an upward concavity at all temperatures and hence with no inflection. Although the 3rd order IMD changes remarkably upon a sudden change in the static magnetic field (not shown), it is relatively insensitive to magnetic relaxation, again in stark contrast to the 2nd order. Thus, although both even and odd order nonlinearity are impacted by the presence of flux, they are not influenced in the same way, indicating different physical processes controlling each order. Although the nonlinear Meissner effect is responsible for enhanced 3rd order nonlinearity near $T_c$[11], and introduction of fluxons at the transmission line edge affects the 3rd order response [25], 2nd order nonlinearity is only possible by breaking of time reversal symmetry and thus should not be expected in the absence of a static magnetic field. Possibly the probe field, which is below 1 MHz, instigates 2nd order nonlinearity when it would not be observed by pure microwaves. This remains a subject of on-going investigation.

## V. CONCLUSION

In summary, even and odd order microwave nonlinearity exhibit fundamentally different behavior in a magnetic field. With two time constants describing the relaxation of 2nd order nonlinearity, it is proposed that 2nd order IMD is influenced by the established mechanisms of remanent magnetic relaxation followed by thermal activation over the surface barrier. Thus 2nd order nonlinearity is correlated to the macroscopic magnetic dipole moment of the superconductor. 3rd order IMD is relatively insensitive to a time effect of magnetic flux highlighting the difference in physical origins of even and odd order nonlinearity with a significant component in the 3rd order nonlinearity that is not related to the macroscopic magnetic dipole moment of the superconductor.